\documentclass[12pt]{article}
\usepackage{graphics}
\input epsf

\newcommand{\pslash}{/\!\!\!\!p}

\newcommand{\be}{\begin{eqnarray}}
\newcommand{\ee}{\end{eqnarray}}

\begin{document}
\title{
\begin{flushright}
{\large UAHEP042}
\end{flushright}
\vskip 1cm
Electron to selectron pair conversion in a susy bubble}
\author{L. Clavelli\footnote{lclavell@bama.ua.edu}\quad and I. 
Perevalova\footnote{perev001@bama.ua.edu}\\
Department of Physics and Astronomy\\
University of Alabama\\
Tuscaloosa AL 35487\\ }
\maketitle
\begin{abstract}
In the standard model, energy release in dense stars is
severely restricted by the Pauli exclusion principle.
However, if, in regions of space of high fermion degeneracy, there 
is a phase transition to a state of exact supersymmetry (SUSY),
fermion to sfermion pair conversion followed by radiative 
transitions to the Bose ground state could lead to a highly collimated 
gamma ray burst.  We calculate the cross section
for $e e \rightarrow {\tilde e}{\tilde e}$ in a SUSY bubble
and construct a monte carlo for the resulting sfermion
amplification by stimulated emission.
\end{abstract}
\renewcommand{\theequation}{\thesection.\arabic{equation}}
\renewcommand{\thesection}{\arabic{section}}
\section{\bf Introduction}
\setcounter{equation}{0}
   Recently it has been proposed that the puzzling gamma ray 
bursts seen over the last few decades in satellite observations 
\cite{Zhang} could be due to a transition
to the SUSY vacuum in compact astrophysical objects such as
white dwarf or neutron stars \cite{CK}.  This phase transition 
could be catalysed by the high matter density of such stars and, 
once nucleated, will spread through the entire star. 
Such a catalysis has been demonstrated in low dimensional models
\cite{Gorsky} at high density and is probably a feature of 
dense matter in arbitrary dimensions \cite{Voloshin}.

In such a SUSY bubble, normal particle pairs can convert to
sparticle pairs, which, being bosons, can drop to the ground
state through gamma ray emission.  In this way, the entire
kinetic energy of a degenerate fermi sea can be radiated 
away.  It is possible that the high degree of collimation
observed in the gamma ray bursts could be due to Bose enhancement 
as is familiar in terrestrial lasers.
If this transition occurs in white dwarf stars, there will be
a multi-component structure to the bursts as various particle
species convert to their SUSY partners, 
possibly interrupted
by periods of fusion of supersymmetric nuclei.  A more
complete discussion of the physical picture proposed is contained 
in \cite{CK}.  This model is dependent on the assumption that the
common particle and sparticle mass in the exact SUSY phase is
equal to (or less than) the particle mass in the broken SUSY phase. 
This mild though necessary assumption is, perhaps, supported by
the string theory result that the ground state supermultiplets are
of low (in fact zero) mass.

     The possibility of a phase transition between vacua requires that
the vacuum structure of the theory is dynamically determined as in
string theory and in certain other models of susy breaking.  
The other possibility, that the susy breaking is determined by 
arbitrary fixed parameters as in the minimal supersymmetric standard
model is probably less satisfying from a theoretical perspective. The
transition we consider (from an unstable de Sitter vacuum with positive
vacuum energy to a stable susy vacuum) has been considered in the formal
string theory study of ref.\cite{KKLT} although phenomenological 
consequences are not part of that work.

 In this article
we treat the simplest component of the phase transition, 
namely the conversion of electron to selectron pairs.
\be
     e^{-}(p_1) e^{-}(p_2) \rightarrow {\tilde e}^{-}(p_3)
   {\tilde e}^{-}(p_4)
\label{eeconv}
\ee
    This cross section was calculated previously in
\cite{Keung} neglecting
the electron mass as appropriate in the broken SUSY phase
where the electron mass is many orders of magnitude less than
the selectron mass.  In the exact SUSY phase the electron and
selectron masses are equal and the cross section near threshold
is needed.  The corresponding simple extensions of
the cross section formulae are given in section II.
The results are applicable to a possible SUSY conversion in
a white dwarf which, in the broken SUSY phase, is stable 
against gravitational collapse due to the electron degeneracy.
In a neutron star, the SUSY conversion would be more complicated
and the energy release would be more slow since sneutrons will
not efficiently radiate.  Even in the white dwarf case, nuclear
processes may be somewhat more important than the electron component
which we treat here.
In section III, we use the cross section of \ref{eeconv} 
in an event generator for the
differential transition rate incorporating the Bose enhancement
effect.  We assume that the entire kinetic energy of the
selectrons is released into gamma radiation as the scalars
drop into the ground state although we do not treat the
radiative processes explicitly in this paper.  Section IV
is reserved for a summary of our conclusions.

\section{\bf Cross section calculation}
\setcounter{equation}{0}

    In the process $e^-e^-\rightarrow\tilde{e}\tilde{e}$ the 
mediator of
the interaction is a Majorana spinor $\tilde{\gamma}$ as shown in
figure 1.

We follow the Feynman rules as reviewed in
\cite{Haber:1984rc}.  In particular the vertex factor for an
incoming left (right) handed Dirac fermion of spinor index $\beta$,
an outgoing SUSY partner, and an outgoing Majorana fermion of
spinor index $\alpha$ is
\be
      -ie \sqrt{2}\quad \frac{(1 \mp \gamma_5)_{\alpha \beta}}{2}
\ee
and for a Majorana fermion line with double incoming arrows as in
figure 1 carrying momentum $q$ from this
vertex to one where the Majorana fermion is emitted with index
$\gamma$, the propagator is
\be
     i \frac{(C^{-1}(\not \! q + m))_{\alpha \gamma}}{q^2-m^2+i\epsilon}.
\ee
Acting on the Dirac spinor the charge conjugation operator, $C$, has 
the effect
\be
       u^{T}(p,s) C^{-1} = - {\overline v}(p,s).
\ee 

For the associated production of left and right selectrons,  
the amplitudes corresponding to graphs $a$ and $b$ in figure 1 are then

\begin{eqnarray}\label{M}
    \nonumber M_a & = &
    \frac{ie^2}{2(t-M_{\tilde{\gamma}}^2)}u^T(p_1)(1+\gamma_5)^T C^{-1}(\pslash_1-\pslash_3+M_{\tilde{\gamma}})(1-\gamma_5)u(p_2)
    \\
    \nonumber M_b & = &
    \frac{ie^2}{2(u-M_{\tilde{\gamma}}^2)}u^T(p_1)(1-\gamma_5)^T C^{-1}({\pslash}_1-{\pslash}_4+M_{\tilde{\gamma}})(1+\gamma_5)u(p_2).
\end{eqnarray}

Here and throughout the paper the Mandelstam variables are

\begin{eqnarray}
 \nonumber s &=& (p_1+p_2)^2 \\
 \nonumber t &=& (p_1-p_3)^2 \\
 \nonumber u &=& (p_1-p_4)^2.
\end{eqnarray}
The squared matrix elements take the form:
\begin{eqnarray}\label{MELR}
   \nonumber \left|\mathcal{M}_{LR}\right|^2 =\Sigma_{ss'} \left( M_aM_a^{\dagger}+M_bM_b^{\dagger}+M_aM_b^{\dagger}+M_bM_a^{\dagger}\right)\\
   \nonumber  = e^4 \left[ \frac{1}{(t-M_{\tilde{\gamma}}^2)^2}\left(ut-2tm_e^2-\left(m_{\tilde{e}}^2-m_e^2\right)^2  \right) \right.\\ +
               \left.     t\leftrightarrow u 
       +\frac{4m_e^2\left(m_{\tilde{e}}^2-m_e^2\right)}{(t-M_{\tilde{\gamma}}^2)(u-M_{\tilde{\gamma}}^2)}\right]
\end{eqnarray}
where $\Sigma_{ss'}$ denotes averaging over the spins of incoming
particles. 
The production of two left or two right selectrons is obtained from the above by the 
appropriate changes in the helicity projection operators and dividing by the statistical
factor for the identical final state particles.

\begin{equation}\label{MELLRR}
  |\mathcal{M}_{RR}|^2=|\mathcal{M}_{LL}|^2=\frac{e^4M_{\tilde{\gamma}}^2}{2!}\left(s-2m_e^2\right)\left( \frac{1}{t-M_{\tilde{\gamma}}^2} + \frac{1}{u-M_{\tilde{\gamma}}^2} \right)^2.
\end{equation}
In the limit of negligible electron mass, these agree with 
the spin-averaged formulae of \cite{Haber:1984rc} and with the squared helicity ampltudes 
of \cite{Keung}.  Other authors have considered the effect of other neutralino exchanges
\cite{Cuypers,Peskin}
but, since these particles in the SUSY phase have masses comparable to the $W$ and $Z$, 
they are negligible for our purposes.

\begin{figure}[tb]
\begin{center}
\epsfxsize= 4.5in 
\leavevmode
\epsfbox{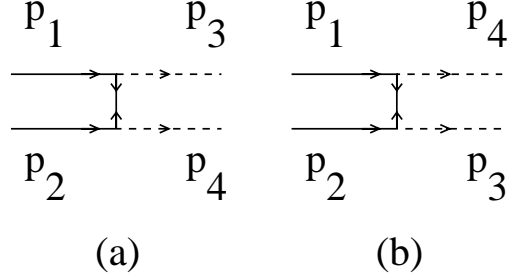}
\end{center}
\caption{Feynman graphs for the conversion of an electron pair
to selectrons, eq.\ref{eeconv}, via photino exchange .}
\label{fig1}
\end{figure}

In the numerical calculations of Section III, we will need the
integrated matrix elements squared: 
\begin{equation}
    f_{AB}(s)=\frac{1}{\sqrt{s(s-4m^2)}}\int_{4m^2-s}^{0}dt|\mathcal{M}_{AB}|^2.
\end{equation}

The total cross sections are related to the $f_{AB}$ by
\be
     \sigma_T(e e \rightarrow {\tilde e}_A {\tilde e}_B) = \frac{f_{AB}(s)}{16 \pi \sqrt{s(s-4 m^2)}}.
\ee

Calculations for the different Matrix elements in the SUSY phase
($m_e=m_{\tilde{e}}=m$) will give us the following:
\begin{eqnarray}
 \nonumber f_{LR}(s) &=& 2 e^4 \sqrt{\frac{s-4m^2}{s}}\left(\frac{-2s -2 M_{\tilde \gamma}^2 + 6 m^2}{s-4m^2+M^2_{\tilde{\gamma}}}+\frac{s-2m^2+2M^2_{\tilde{\gamma}}}{s-4m^2}\ln{\frac{s-4m^2+M^2_{\tilde{\gamma}}}{M^2_{\tilde{\gamma}}}}\right) \\
 \nonumber f_{LL}(s) &=&f_{RR}(s)=\frac{4 e^4 (s-2m^2)}{\sqrt{s(s-4m^2)}}\left(\frac{s-4m^2}{(s-4m^2+M^2_{\tilde{\gamma}})} + \frac{2 M^2_{\tilde{\gamma}}}{s-4m^2+2M^2_{\tilde{\gamma}}}\ln{\frac{s-4m^2+M^2_{\tilde{\gamma}}}{M^2_{\tilde{\gamma}}}}\right).
\end{eqnarray}

\begin{figure}[tb]
\begin{center}
\epsfxsize= 4.5in 
\leavevmode
\epsfbox{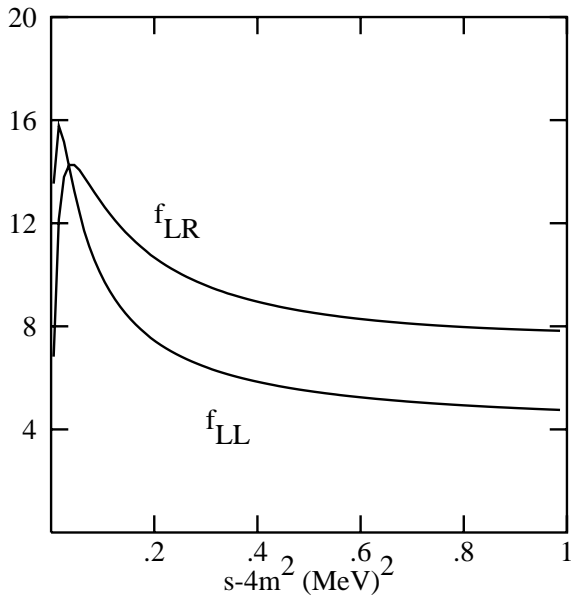}
\end{center}
\caption{Behavior of the $t$ integrated matrix elements squared as a
function of $s-4 m^2$.}
\label{fig2}
\end{figure}

These are only logarithmically sensitive to the photino mass.  
The selectron momentum and
angular distributions to be calculated in the next section are even less sensitive to the
photino mass since they depend only on the shape of the cross sections and not on the
absolute values.  Integrals of $f_{LL}$ and $f_{LR}$ as required in the conversion rates
discussed in the next section are finite in the limit of zero photino mass.  
However, a more precise treatment of photino mass effects requires a
careful treatment of experimental resolution beyond the scope of this study. 
The behavior of the functions
$f_{LL}(s)$ and $f_{LR}(s)$ for $M_{\tilde{\gamma}}= m/4$ is shown in figure 2. For larger
values of the photino cutoff mass, $f_{LR}$ lies below $f_{LL}$ although the near constant
values of the two $f's$ well above threshold are not affected.

\section{\bf Event generation}
\setcounter{equation}{0}

     In a typical white dwarf (solar mass and earth radius)
there are 
\be
       N_0 = 6 \cdot 10^{56}
\ee
electrons in a degenerate Fermi sea:
\be
      dN = \frac{2 p^2 dp \quad d(cos(\theta)) \quad d\phi}{(2 \pi \hbar)^3} .
\label{sea}
\ee
The Fermi momentum is
\be
      p_{max}=\left (\frac{3 N}{8 \pi V} \right)^{1/3} 2 \pi \hbar
 = 0.498 (\frac{n}{n_0})^{1/3} MeV/c   .
\label{pmax}
\ee
Here $n$ is the electron number density and $n_0$ is that of the
nominal white dwarf of solar mass and earth radius.  
The event rate for process \ref{eeconv} in a volume V
is given in terms of the differential cross section by
\be
    \Gamma = \int d\sigma \left( \frac{dN_i v_i}{V} 
 \right) dN_t  .
\ee
where the incident and target distributions are given by \ref{sea}.
The differential cross section is given in terms of the invariant
matrix element squared by
\be
     d\sigma = \frac{{\mid \cal{M} \mid}^2 d\Omega_f}{4 E_1 E_2 v}.
\ee
Here $m$ is the common electron and selectron mass in the SUSY
bubble.  
The energies, $E_i$ and the incident velocity, $v$, are those of the
target rest frame.
The final state phase space is
\be
     d\Omega_f = \frac{d^3p_3}{2E_3}\frac{d^3p_4}{2E_4}\frac{\delta^4(p_1+p_2-p_3-p_4)}
     {(2 \pi)^2}  
\ee
where any statistical factors that occur for identical final state particles are put
into the matrix element squared.
Thus the Lorentz invariant event rate per unit volume is
\be
      \frac{d\Gamma}{V} = \frac{4 {\mid \cal{M} \mid}^2}{(2 \pi)^8}
       \frac{d^3p_1}{2E_1}\frac{d^3p_2}{2E_2}\frac{d^3p_3}{2E_3}\frac{d^3p_4}{2E_4}
      \delta^4(p_1+p_2-p_3-p_4)  .
\ee
The $p_2$ integral can be done trivially using the $\delta$ function and the $p_1$
integral is then conveniently done in the CM frame with $\hat{p}_3 = \hat{e}_z$.
\be
   \frac{d^3p_1}{2E_1} \delta((p_3+p_4-p_1)^2-m^2) = \frac{\pi dt}{2 \sqrt{s(s-4m^2)}} .
\ee
Thus
\be
      \frac{d\Gamma}{V} = \frac{{\mid \cal{M} \mid}^2}{(2 \pi)^7 \sqrt{s(s-4m^2)}}
        dt \frac{d^3p_3}{2E_3}\frac{d^3p_4}{2E_4}  .
\ee
or
\be
      \frac{d\Gamma}{V} = \frac{{\mid \cal{M} \mid}^2}{(2 \pi)^7 \sqrt{s(s-4m^2)}}
        dt \frac{{p_3}^2}{2E_3}\frac{{p_4}^2}{2E_4}dp_3 dcos(\theta_3)d\phi_3
        dp_4 dcos(\theta_4)d\phi_4  .
\label{dG}
\ee
For $p_{max}$ we use the value of \ref{pmax} corresponding to the Fermi energy 
in a white dwarf of earth radius and solar mass.

    The final state of the process consists of two distinct species of scalars,  
${\tilde e}_L$ and ${\tilde e}_R$. Thus the effective matrix element 
squared in \ref{dG}
and elsewhere above is actually
\be
     {\mid {\cal M}\mid ^2} = {\mid {\cal M}_{LR}\mid ^2}+ 
       ({\mid {\cal M}_{LL}\mid ^2} +{\mid {\cal M}_{RR}\mid ^2}) .
\ee 
In a bath of pre-existing selectrons each matrix element squared is related to the
elementary matrix element squared calculated with no pre-existing selectrons by the
Bose statistical factors
\be
\nonumber
     {\mid {\cal M}\mid ^2} &=& {\mid {\cal M}_0}_{LR}\mid ^2 (N_L(\vec{p}_3)+1)
       (N_R(\vec{p}_4)+1)+ 
   ({\mid {\cal M}_0}_{LL}\mid^2 (N_L(\vec{p}_3)+1)(N_L(\vec{p}_4)+1) \\
         &+&{\mid {\cal M}_0}_{RR}\mid ^2 (N_R(\vec{p}_3)+1)(N_R(\vec{p}_4)+1)) .
\label{M2}
\ee 
The matrix elements of the previous section, calculated for the case of no selectrons
in the initial state, are the ${\cal M}_0$ of this section.
For the present we treat only the $LL$ final
state.  The $RR$ final state implies that there could be at least four gamma ray jets
in each burst and possibly many more if 
different energy levels in the Fermi sea lead to independent gamma ray bursts.
This suggests a picture where the gamma ray jets are much more numerous in each
stellar explosion
and, individually, much more narrow and less energetic than currently assumed.
A more detailed analysis, beyond the scope of the current paper, is needed to
explore this point but we speculate that this could be the cause of the "spikey"
nature or rapid time variability of many of the observed bursts. 

     Our remaining analysis in this paper, restricted to the $LL$ jets, allows us
to drop, the $L$ subscript on the occupation numbers.
The CM energy $\sqrt{s}$ is determined by $\vec{p}_3$ and $\vec{p}_4$.  The $t$ integral
has been done analytically in section 2:

   We define a grid of $N_{bin}$ points
in each of the six variables.  In practice we choose $N_{bin}=15$.
The discretized $i'th$ integration variable in \ref{dG} is
\be
      v_i = v_{i,min}+(v_{i,max}-v_{i,min})(N_i+1/2)/N_{bin}
\ee
where
\be
       0 \leq N_i \leq N_{bin}-1 .
\ee
In order to generate events with probability defined by \ref{dG} and \ref{M2}, it is
convenient to linearize the six dimensional integral.  We define the composite integer
variable
\be
      j = \sum_{i=1}^6 N_{bin}^{6-i}N_i
\ee
with limits
\be
      0 \leq j \leq N_{bin}^6 -1.
\ee
Each value of $j$ corresponds to a unique value of each of the six integration variables.
The distribution of events is then defined as an integer valued array with $N_{bin}^6$
grid points.  To handle $15^6$ grid points requires careful memory management techniques.
\footnote{We thank Doug Leonard for consultation on these techniques.}
The integer $j$ can be decomposed into two integers $j_3$ and $j_4$ which encode the
three dimensional vectors $\vec{p_3}$ and $\vec{p_4}$ respectively.
\be
\nonumber
             j_3 &=& j / N_{bin}^3 \\
             j_4 &=& j \bmod N_{bin}^3 \\
\nonumber
             j &=& N_{bin}^3 j_3 + j_4  .
\ee
The first of these equations is defined by integer division, i.e. $j_3$ is the largest
integer less than or equal to $j/N_{bin}^3$.  The event generation follows standard
techniques \cite{PDG} except that the probability distribution changes with each event 
due to the Bose enhancement factors.  In the event generation of the selectron 
distributions, constant overall factors
in \ref{dG} play no role.
We begin by putting all the selectron occupation numbers to zero and calculating the (unnormalized) probabilities
\be
      P(j) = f_{LL}(s) \frac{p_3^2 p_4^2}{E_3 E_4} (N(j_3)+1)(N(j_4)+1)  .
\ee       
We then define the partial sum
\be
      R(j) = \sum_{i=0}^j P(i)
\ee
as well as the complete sum
\be
      P_{int} = \sum_{i=0}^{N_{bin}^6-1}P(i) .
\ee
$R(j)$ is a monotonically increasing function of j.  One then calls a random number 
$w$ between zero and
one.  For the unique value of $j$ for which $w > R(j-1)/P_{int}$ and 
$w \leq R(j)/P_{int}$ one increments the selectron occupation numbers by one:
\be
\nonumber
      N(j_3) \rightarrow N(j_3)+ 1 ,\\
      N(j_4) \rightarrow N(j_4)+ 1 .
\ee
$P(j\prime)$ changes in the $j th$ bin only by a factor
\be
      f = (N(j_3)+1)(N(j_4)+1)/(N(j_3)N(j_4)).
\label{f}
\ee
Here $N(j_3)$ and $N(j_4)$ are the new occupation numbers (after incrementing).
$R(j\prime)$ changes for each $j\prime\geq j$
\be
      R(j\prime) \rightarrow R(j\prime) + P(j)(f-1)  \qquad{\textstyle for}\qquad j\prime \geq j .
\ee
In addition
\be
     P_{int} \rightarrow P_{int}+P(j)(f-1) .
\ee  
After making these replacements, one adjusts $P(j)$:
\be
     P(j) \rightarrow P(j) f  .
\ee
One then repeats the process as many times as possible choosing new random values of $w$.
After very many events the distribution is amplified at two particular values of $\vec{p_3}$
and $\vec{p_4}$.  

\begin{figure}[htb]
\begin{center}
\epsfxsize= 4.5in 
\leavevmode
\epsfbox{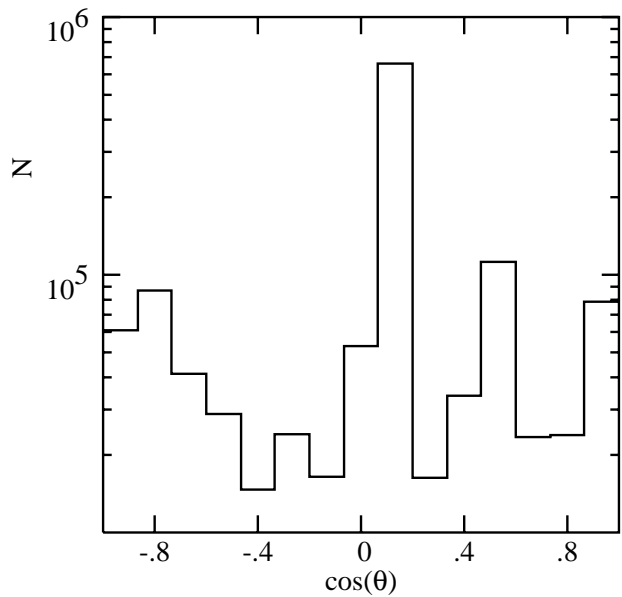}
\caption{angular distribution after $600,000$ throws, showing 
effect of Bose enhancement.  In this run, the bins are incremented
twice at each throw}
\label{angdist}
\end{center}
\end{figure}

     In the simulation two problems arise.  The first is that the enhanced
probability to put subsequent events in some particular $j'th$ bin is a very mild 
enhancement at first.  Only after some huge number of events does the structure lock in
on a particular $j$ value.  Since the number of available fermions in a compact star
is of order $10^{56}$, this is not a problem in principle but it does pose technical
problems with computers of currently available speed and memory.  
To accelerate the buildup of jet
structure, instead of incrementing the selectron occupation numbers by one at each
throw of the dice, we increment by two.  If one increments by more than one at each throw, 
the $f$ of \ref{f}  
has the obvious redefinition.   The distributions in selectron momentum, polar angle
cosine, and azimuthal angle after $600,000$ throws (or $1.2$ million events) 
is shown in table I.  The jagged polar angular distribution shown in figure \ref{angdist}
is an interesting feature of the calculation.

     In a simplified run, (not shown here) where we look at the conversion
of two electrons at the Fermi surface, although there are no observable jets in the
first million events, a clear jet structure emerges at $9 \cdot 10^{7}$ events even
when selectron occupation numbers are incremented by one at each event as is physically
required.  In this run it is not required that the electron pair has zero total 
momentum in the rest frame of the star.

\begin{table}[ht]
\begin{center}
\begin{tabular}{|rr|rr|rr|}\hline
 p (MeV)  &   N(p)   & $\cos \theta$  & N($\cos \theta$)  &   $\phi$   & N($\phi$)\\
\hline
 0.017 	  &    42    &	 -0.93 	 &   60914   &	 0.209  &	   16692 \\  
 0.050 	  &   360    &	 -0.80 	 &  87034    &	 0.628  &	   56060 \\
 0.083 	  &   986    &	 -0.67 	 &  41284    & 	 1.047  &	   94394 \\
 0.116 	  &  1948    &	 -0.53 	 &  28832    &	 1.466  &	   30174 \\
 0.149 	  &  3178    &	 -0.40 	 &  14672    &	 1.885  &	   63792 \\ 
 0.183 	  &  4690    &	 -0.27 	 &  24064    &	 2.304  &	  127048 \\ 
 0.216 	  &  6572    &	 -0.13 	 &  16452    &	 2.723  &	  607026 \\ 
 0.249 	  &  8410    &	 0.00 	 &  52874    &	 3.142  &	   13538 \\ 
 0.282 	  & 11510    &	 0.13 	 & 660320    &	 3.560  &	   50778 \\ 
 0.315 	  & 13548    &	 0.27 	 &  16302    &	 3.979  &	   21316 \\ 
 0.349 	  & 17198    &	 0.40 	 &  33942    &	 4.398  &	   46504 \\ 
 0.382 	  & 26284    &	 0.53 	 & 112400    &	 4.817  &	   25394 \\ 
 0.415 	  & 63298    &	 0.67 	 &  23458    &	 5.236  &	   13670 \\ 
 0.448 	  & 82526    &	 0.80 	 &  23878    &	 5.655  &	   31868 \\ 
 0.481 	  & 1034494  &	 0.93 	 &  78618    &	 6.074  &	   76790 \\ 
\hline
\end{tabular}
\end{center}
\caption{Selectron momentum and angular distributions showing the effect of
boson enhancement.  In this run the occupation numbers are incremented by two
at each throw of the dice.}
\end{table}

\begin{figure}[htb]
\begin{center}
\epsfxsize= 4.5in 
\leavevmode
\epsfbox{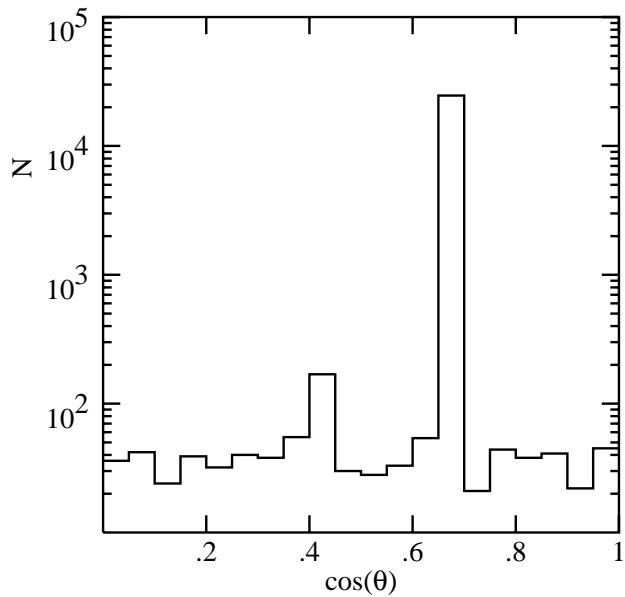}
\caption{Polar angle distribution for conversion of those
electron pairs whose center of momentum is that of the star.  One quadrant
of the angular space is shown with a balancing selectron jet in the opposite
direction.}
\label{histo}
\end{center}
\end{figure}

      The second problem is the following.  We have treated the star as having an 
inexhaustible
supply of electrons of each momentum in the fermi sea.  In practice this number is very
large but not infinite.  Only after all the electrons have been exhausted will the 
selectron momenta add to zero in the rest frame of the star.  For the present one
can artificially circumvent
this problem by considering electron to selectron conversion among those electron pairs
whose center of momentum frame is that of the star.  Then we can choose one selectron
according to the distribution of \ref{dG} with the other selectron necessarily having 
the balancing momentum.  In this case we can rapidly see the jet structure develop 
incrementing each selectron occupation number by one at each throw of the dice. 
The corresponding selectron distribution is shown in figure \ref{histo} after only
$50,000$ events.

\section{\bf Conclusions}
\setcounter{equation}{0}

    The results after $600,000$ throws of the dice are shown in table 1.  
Since the integrated squared matrix elements vanish at threshold and the
phase space element favors high momenta,
the typical selectron momentum is thus
somewhat greater than the estimate in \cite{CK} based on the average energy in 
the electron Fermi sea.  The selectron momentum distributions given in table 1 
will more or less directly carry over into photon distributions since bosons will
necessarily fall into the ground state via photon emission.  Thus the firm prediction
of $MeV$ level photons in the burst is a quantitative, and perhaps the primary, 
result of the current paper.  A secondary result is that, due to the increased 
production probability for higher photon energy (i.e. selectron kinetic energy in
table I),
the initial spikes in a gamma ray burst would be expected to be of higher average
photon energy than later components.  This is consistent with observations \cite{Pian}  

The minimum duration of the burst was fixed in \cite{CK} by 
assuming that the SUSY bubble, once nucleated, would grow at the speed of light and
that photons in the dense star would travel with effective index of refraction one.
Both of these assumptions might need to be more closely examined. 

In particular, since one should probably regard the bubble surface as a mechanical membrane,
a more reasonable estimate of the burst duration might be given by assuming that
the bubble expands at the speed of sound rather than the speed of light.  Then, if one
takes into account the range of masses and radii among white dwarfs and if one estimates
the bubble expansion speed by the speed of sound in a star of corresponding average density,
the range of predicted burst durations extends well into that of the observations.  A detailed
study of the duration distribution taking into account density inhomogeneities and other effects
is a high priority subject of future study. 

    Finally, we should discuss our expectation that the susy bubble is confined 
to the dense star and does not escape to take over the universe.
If one is in the false vacuum of broken susy,
bubbles of true vacuum (exact susy) are constantly being
produced with a steeply falling distribution in bubble radius, r.
At creation, or at any later stage in its development, a bubble of radius r 
will expand or contract depending on which behavior is energetically favorable.
The condition for expansion depends on the surface tension, S, of the bubble
and the energy density in the region immediately outside the bubble.
Consider, for example, a bubble of exact susy in a larger region of broken
susy.  Outside the bubble the energy density will be $\epsilon + \rho$
where $\epsilon$ is the vacuum energy  and $\rho$ is the outside ground state
matter density if any.  If the susy bubble were to make a virtual expansion
into an infinitesimally larger spherical shell, its ground state 
energy density would
be $\rho_s$.  The difference between $\rho$ and $\rho_s$ is the excitation
energy density of the electrons in the broken susy phase.
Classically, the ground state energy after such a virtual
expansion minus the previous ground state energy is
\be
    \Delta E = \frac{4 \pi}{3}\left( (r+\delta r)^3 - r^3\right) (\rho_s - \rho - \epsilon)
                 + 4 \pi S \left( (r+\delta r)^2 - r^2 \right)
\ee
or
\be
    \Delta E = - 4 \pi r \delta r \left( r (\epsilon + \Delta \rho) - 2 S \right)  
\label{deltaE}
\ee
where we have put 
\be
    \Delta \rho = \rho - \rho_s .
\ee

Classically therefore, the system will find it energetically advantageous to expand if
$r > \frac{2S}{\epsilon + \Delta \rho}$.  Similarly, the bubble will contract if its
radius is below this density dependent critical value.   A more exact instanton calculation \cite{Coleman}
in vacuo ($\rho = \rho_s = 0$) replaces $2S$ by $3S$ in eq.\ref{deltaE}.  One would, therefore, expect the critical radius for a susy bubble to be
\be
    R_c = \frac{3S}{\epsilon + \Delta \rho}
\ee
In vacuo or ignoring the effect of the Pauli principle, $\Delta \rho=0$.  In a 
homogeneous region, if a bubble is created at greater than the critical radius, it will expand indefinitely.  If however, the bubble comes to the boundary of a 
dense region outside of which $\rho$ and $\Delta \rho$ are zero, the critical
radius jumps discontinuously to its vacuum value, effectively confining the bubble to the high density region.

    Given the indirect hints of supersymmetry from dark matter and accelerator experiments,
given the positive vacuum energy suggested by the acceleration of the universe, and
given the persistent suggestion of string theory that the true vacuum of the theory is
supersymmetric, a phase transition from our broken-SUSY phase to the exact SUSY phase is
probably inevitable.  There are also good reasons to believe the transition to the true vacuum
would be catalyzed in dense matter \cite{Voloshin}.  In addition, given the Bose nature
of the final state particles, collimation due to stimulated emission is to be expected
on physical grounds.  Only our assumption that the
probability of transition is sufficient in compact stars to reproduce the rate of gamma
ray bursts might be considered speculative.  

    It is, however, clear that many points remain to be investigated in the current 
SUSY phase transition picture of gamma ray bursts.  Some of these are:

\begin{enumerate}
\item {\bf calculation of the hadronic component of SUSY conversion.} This will become 
 even more essential if the bursts originate in Neutron stars and not in white dwarfs.
\item {\bf calculation of the gamma ray spectrum} from radiative SUSY conversion and
 (not independent) bremstrahlung from the final state selectron gas.  Since the selectrons
 are bosonic, all of their kinetic energy will emerge as photons. 
\item {\bf depletion effects in the Fermi sea}
\item {\bf fusion of SUSY (or partially SUSY) nuclei.} This could lead to temporary
interruption of the gamma ray bursts.
\item {\bf polarization effects}
\end{enumerate}

    Only after many of these effects have been studied can one attempt to predict the
angular width of jets, the energy spectra, the "light curves", and other features of
the rapidly growing observational data.  As shown here and in \cite{CK} the SUSY phase
transition model provides a framework for discussing other details of this phenomenon
that is alternative to the more traditional astrophysical approaches.  These latter
approaches based on relativistic ejection of large bodies of neutral matter from
black hole accretion disks with the subsequent conversion of a large amount of 
kinetic energy into collimated gamma rays on sub-second time scales attempt to
describe gamma ray bursts within
the boundaries of standard model physics but have not as yet
led to sharp predictions for typical photon energies or total burst energies 
untied to free parameters in the models.  In addition the physical basis of the energy 
release
(central engine) or the mechanism for production of collimated gamma rays are not yet
as fully defined in the accretion models as in the current phase transition model.
For recent papers marking the
current status of traditional astrophysical approaches and giving references to
earlier work along those lines see \cite{Lee} and \cite{Yamazaki}.

    Another important area that needs study is the possible role of a SUSY phase 
transition in supernova collapse.  This is potentially of great interest since it has
become apparent in recent years that the current standard model of supernova explosion
(energy deposited by neutrinos in a surrounding shell of matter) is not effective in
producing the observed explosions \cite{Duan,Bethe}.  This problem might also raise
questions about the efficiency of $\nu \overline{\nu}$ annihilation into an $e^+ e^-$
cloud available for production of a relativistic fireball to explain gamma ray bursts
as suggested in \cite{Lee}.

      Examples of other new physics suggestions for gamma ray 
bursts can be found in 
\cite{Sannino} and references contained therein.  Again, it seems that these models might
not have the same success in predicting the salient features of gamma ray bursts as does
the SUSY phase transition model.

{\bf Acknowledgements}

    This work was supported in part by the US Department of Energy under grant 
DE-FG02-96ER-40967. We gratefully acknowledge discussions with Doug Leonard, 
Phil Hardee, Bill Keel, George Karatheodoris, and Wai-Yee Keung.

\par
\end{document}